\documentstyle[12pt]{article}
\def\kk{\hbox{K\kern-0.59em\raise0.4ex\hbox{$\scriptstyle\bf |$}}}
\def\cc{\hbox{C\kern-0.55em\raise0.4ex\hbox{$\scriptstyle
 |$}}}
\newcommand{\dotA}{\stackrel{\ \bf .}{\bar{A}}}
\newcommand{\dotalpha}{\stackrel{\bf .}{\bar{\alpha}}}
\newcommand{\dotz}{\stackrel{\bf .}{\bar{z}}}
\newcommand{\dotxi}{\stackrel{\ \bf .}{\bar{\xi}}}
\newcommand{\be}{\begin{eqnarray}}
\newcommand{\ee}{\end{eqnarray}}
\newcommand{\add}{\addtocounter{equation}{-1}}

\begin{document}
\hfill FTUV/93-7
\sl

\begin{center}
{\bf  PATH INTEGRALS FOR QUANTUM ALGEBRAS\\
AND THE CLASSICAL LIMIT$^*$}
\vskip 0.5 cm
{\bf  Demosthenes Ellinas}\\
Departamento de Fisica Teorica  Facultad de Fisica \\
Universidad de Valencia  E-46100 Burjasot Valencia Spain\\
{\it ellinas@evalvx.ific.uv.es}
\vskip 2.0 cm
\end{center}
\begin{abstract}
{\sl Coherent states path integral formalism for the
simplest quantum algebras, $q$-oscillator, $SU_q(2)$ and $SU_q(1,1)$
is introduced. In the classical limit canonical
structure is derived with modified symplectic and Riemannian metric.
Non-constant deformation induced curvature for the phase spaces is
obtained.}
\end{abstract}

\pagebreak
\vskip 1.0 cm

 Aiming to a better understanding of the notion of quantum groups and
algebras [1-3], is desirable to try to elucidate the geometrical properties
of these structure. The analogues properties of Lie groups emerge from the
study of those groups as transformations groups in some spaces. The
symplectic and Riemannian properties of classical groups are well known
and is the purpose of this paper to attempt to set similar questions
when $q$-deformation is in presence. Confronted with the vast generality of
such
a problem we have done here the minor choice to study three of the simplest
quantum algebras, namely the deformed Weyl-Heisenberg algebra $(q$-WH) [4,5]
the $SU_q(2)$ and $SU_q(1,1)$ algebras. To this end we will employ the
tools of path integrals and coherent states, adapted to the $q$-deformed
situation.

Usual coherent states (CS) [6,7] for simple Lie groups are widely used in the
path
integral formalism  especially for Hamiltonians which are elements of the
corresponding Lie algebra of the groups. The same formalism also provides in
the
classical limit, regarded as the case  where the Planck constant is
taken small compared to
the action, the canonical equations of motion in the phase spaces which for
the groups we are here concerned are the coset spaces $\frac{WH}{U(1)}\approx
{\bf R}^2$, $\frac{SU(2)}{U(1)}\approx S^2$ and $\frac{SU(1,1)}{U(1)}\approx
S^{1,1}$. Obviously in the cases of $SU(2)$ and $SU(1,1)$ groups the ensuing
coset spaces i.e. sphere and hyperboloid are generalizations of the usual plane
phase space of the harmonic oscillator and this gives rise to a K\"ahler
manifold structure with modified canonical symplectic 2-form [8].

The deformed CS [5,9] which will be used here to built up the propagator
satisfy the completeness relation and are obtained by acting on some lowest
weight with a displacement operator which involves the deformed or
alternatively the ordinary exponential [10,9]. This second possibility,
as will be seen in the following, has some important technical merits
in the constraction of path integrals and in the study of the eigenvalue
problem of the $q$-CS.

A brief of our results is as follows: in a very similar manner to the changes
of the geometrical structure of the harmonic oscillator phase space
occurring when we pass to the $SU(2)$ and $SU(1,1)$ phase spaces,
the $q$-deformation of the above algebras modifies their corresponding
phase spaces, as is shown by evaluating the  symplectic and Riemann
metrics. Also the computation of the curvature scalar reveals that a
$q$-deformation induces a non-constant curvature in each of the
above phase spaces.

\vskip 1.0 cm
 Deformed coherent states for the $q$-WH algebra are defined by
$(\alpha\in\cc \ )$
\be
|\alpha)_q=e^{\alpha a^+_q}_q|0>=e^{\alpha T^+_a}|0>=\sum\limits^\infty
_{n=0}\frac{\alpha^n}{\sqrt{[n]!}}\ |n>
\ee
where the $q$-oscillator commutation relations are:
\be
a_qa^+_q-a^+_qa_q=[N+1]-[N]\hspace{0.5cm}{\rm ,}\hspace{0.5cm}[N,a^+_q]=a^+_q
\hspace{0.5cm}{\rm and}\hspace{0.5cm}[N,a_q]=-a_q
\ee
while
\be
T^+_a=a^+_q\frac{(N+1)}{[N+1]}\hspace{1.5cm}{\rm and}\hspace{1.5cm}
T^-_a=\frac{(N+1)}{[N+1]}a_q
\ee
and the following symbol are used $(q=e^\gamma)$:
$$[n]=\frac{q^n-q^n}{q-q^{-1}}\hspace{1.5cm}{\rm and}\hspace{1.5cm}
[n]!=[1][2]\cdots[n]$$
with the $q$-exponential function $e^x_q=\sum\limits^\infty_{n=0}
\frac{x^n}{[n]!}$ [11].
The normalized states
$|\alpha>_q=\frac{1}{\sqrt{_q\alpha|\alpha)_q}}|\alpha)_q$
with $_q(\alpha|\alpha)_q=e_q^{|\alpha|^2}$ are eigenstates of the annihilation
operator $a_q|\alpha>_q=\alpha|\alpha>_q$ and satisfy the completeness
relation [12-14]
\be
{\bf 1}=\int|\alpha>_q d\mu_q(\alpha)_q<\alpha|\hspace{1.5cm}{\rm with}
\hspace{1.5cm}d\mu_q(\alpha)=\frac{d^2_q\alpha}{ _q(\alpha|\alpha)_q}
\ee
where the integral is regarded as the Jackson's $q$-integral [15]. These $q$-CS
are minimum-uncertainty states in the sence that they minimize the
$[q_q,p_q]$ commutator
$$\Delta q_q\Delta p_q=\frac{1}{2}|_q<\alpha|[q_q,p_a]|\alpha>_q|$$
where $a_q=\frac{1}{\sqrt{2}}(q_q+ip_q)$ and $a^+_q=\frac{1}{\sqrt{2}}
(q_q-ip_a)$.
\vskip 1.0 cm
 The deformed CS for the $SU_q(2)$ algebra, related to representations
characterized by j=1/2, 1, 3/2,... are defined by
$(z\in\cc \ )$
\be
|z)=e^{zJ^+_q }_q|-j>=e^{zT^+_J}|-j>=\sum\limits^j_{m=-j}\left[\begin{array}
{c}2j\\j+m\end{array}\right]_qz^{j+m}|m>
\ee
where the $q$-binomal is defined as $[^a_b]_q=\frac{[a]!}{[b]![a-b]!}.$
The generators involved in the definition satisfy the commutation relations
\be
[J^3_q,J^\pm_q]=\pm J^\pm _q\hspace{1.5cm}[J^+_q,J^-_q]=[2J^3_q]
\ee
while
\be
T^+_J=J^+_q\frac{(J^3_q+j+1)}{[J^3_q+j+1]}\hspace{1.5cm}{\rm and}
\hspace{1.5cm}T^-_J=\frac{(J^3_q+j+1)}{[J^3_q+j+1]}J^-_q
\ee
The factor $(1+|z|^2)^{2j}_q\equiv\ _q(z|z)_q$ normalizes the states,
$|z>_q=\frac{1}{\sqrt{_q(z|z)_q}}|z)_q$ and using the general formula
$$(x+y)^n_q\equiv\sum\limits^n_{m=0}\left[\begin{array}{c}n\\m\end{array}
\right]_qx^{n-m}y^m=\prod\limits^n_{k=1}(x+q^{n-2k+1}y)$$
derived with the help of $[2m+1]=\sum\limits^m_{\ell=-m}q^{2\ell}$,
is written as
$$_q(z|z)_q=\sum\limits^j_{m=-j}\left[\begin{array}{c}2j\\m+j\end{array}
\right]_q\ |z|^m=\prod\limits^{2j}_{k=1}(1+q^{2j-2k+1}|z|^2)\ .$$
The normalized $q$-CS are complete with resolution of unity
$${\bf 1}=\int|z>_qd\mu_q(z)_q<z|\hspace{1.5cm}{\rm with}\hspace{1.5cm}
d\mu_q(z)=\frac{[2j+1]}{_q(z|z)^2_q}d^2_qz\ ,$$
where again the Jackson's $q$-integral is in use.

It is also interesting that the $q$-CS satisfy the eigenvalue problem,
\be
(J^-_q+(q^j+q^{-j})z[J^3_q]-z^2J^+_q)|z>_q=0
\ee
which upon taking the zero deformation limit reduces to its $q=1$
analogue equation which serves as a definition, up to a phase factor, for the
$SU(2)$ coherest state [7]. For future use we also record the formula
\be
J^\pm_q|z>_q=z^{\mp 1}[j\pm J^3_q]|z>_q\ .
\ee
\vskip 1.0 cm
 The coherent states related to the quantum $SU(1,1)$
\be
[K^3_q,K^\pm_q]=\pm K^\pm_q\hspace{1.5cm}[K^+_q,K^-_q]=-[2K^3_q]\ ,
\ee
and associated with the discrete representations characterized by
k=1, 3/2, 2, 5/2, ...
are defined by the generators
\be
T^+_K=K^+_q\frac{(K^3_q-k+1)}{[K^3_q-k+1]}\hspace{1.5cm}{\rm and}
\hspace{1.5cm}T^-_K=\frac{(K^3-k+1)}{[K^3_q-k+1]}K^-_q\ ,
\ee
in a manner similar to the previous cases:
\be
|\xi)_q=e_q^{\xi K^+_q}|k;0>=e^{\xi T^+_K}|k;0>=\sum\limits^\infty
_{n=0}\frac{[2k+n+1]!}{[n]![2k+1]!}\xi^n|k;n>\ ,
\ee
where $\xi\in D^k=\{|\xi|^2<q^{k-1}\}$.
With normalization factor obtained from the overlap of two states
\be
(1-|\xi|^2)^{-2k}_q\equiv\ _q(\xi|\xi)_q=\sum\limits^\infty_{n=0}
\frac{[2k+n+1]!}{[n]![2k+1]!}|\xi|^{2n}
\ee
the normalized states are complete,
$${\bf 1}=\int|\xi>_qdu_q(\xi)_q<\xi|\hspace{1.5cm}{\rm with}
\hspace{1.5cm}du_q(\xi)=\frac{[2k-1]}{_q(\xi|\xi)^{-2}_q} d^2_q\xi\ ,$$
and obey the equations,
$$(K^-_q+(q^k+q^{-k})\xi[K^3_q]+\xi^2K^+_q)|\xi>_q=0$$
and
$$K^\pm_q|\xi>_q=\xi^{\mp 1}[K^3_q\mp k]|\xi>_q\ .$$
\vskip 1.0 cm
 We proceed now with the $q$-CS propagator utilizing the
coupleteness relations of the $q$-CS. Let $(A=\alpha,z,\xi)$,
the transition amplitude between coherent states takes the form
\be
\kk =<A''|U(t'',t')|A'>=\int{\cal D}\mu_q(A)\exp\biggl[\sum\limits^L_
{\ell=1}\ell n<A_\ell|A_{\ell-1}>-\frac{i}{\hbar}\varepsilon\frac{<A_\ell
|H|A_{\ell-1}>}{<A_\ell|A_{\ell-1}>}\biggr]
\ee
where
$${\cal D}\mu_q(A)=\lim\limits_{^{L\rightarrow\infty}_{\varepsilon
\rightarrow 0}}\prod\limits^{L-1}_{\ell=1}d\mu_q(A_\ell)\ .$$
and $\varepsilon=\frac{t''-t'}{L}$, while $H$ stands for the Hamiltonian
which explicit form is not important.
In the classical limit, considered here as the case where $\hbar\ll$ action,
while
the deformation parameter $q$ is retained fixed, assuming that $A_{\ell-1}\cong
A_
\ell-\Delta A_\ell$, then from the definition of the CSs and the short-time
 approximation follows that
\be
\varepsilon.\frac{1}{\varepsilon}\ell n<A_\ell|A_{\ell-1}>\cong
\frac{\varepsilon}{2}(\frac{\Delta\bar{A}_\ell}{\varepsilon}<A_\ell|
T^-_i|A_\ell>-\frac{\Delta A_\ell}{\varepsilon}<A_\ell|T^+_i|A_\ell>)
\ee
where $i=a,J,K$ and bar denotes complex conjugation.In the limit where
$L\rightarrow \infty $ and $\varepsilon\rightarrow 0$ the r.h.s of the
above expression is written formally as $\frac{1}{2}(\dot{\bar A}<A|T^+_i|A>
-\dot{A}<A|T^-_i|A>)dt$, where overdot denotes time derivative.
In that limit
$$\kk=\int{\cal D}\mu_q(A)\exp[\frac{i}{\hbar}\int\limits
^{t''-t'}_0\ dt\ L(A,\bar{A};\dot{A}\dotA)]$$
and the Langrangian is given by,
\be
L=\frac{i\hbar}{2}[\dot{A}<A|T^+_i|A>-\dotA<A|T^-_i|A>]-{\cal H}
(A,\bar{A})\ ,
\ee
where ${\cal H}=<A|H|A>$, from which we extract the canonical 1-form
\be
Q=\frac{i\hbar}{2}(<A|T^+_i|A> dA-<A|T^-_i|A> d\bar{A}) .
\ee
By using the properties of the $q$-CS as above we obtain for the three
cases $(<\cdot>\equiv <A|\cdot|A>)$
\renewcommand{\theequation}{\arabic{equation}a}
\be
L=\frac{i\hbar}{2}\biggl<\frac{N+1}{[N+1]}\biggr>(\dot{\alpha}\bar{\alpha}
-\dotalpha\alpha)-{\cal H}(\alpha,\bar{\alpha})\ ,
\ee
and
\add
\renewcommand{\theequation}{\arabic{equation}b}
\be
L=\frac{i\hbar}{2}<J^3_q+j>(\dot{z}z^{-1}-\dotz\bar{z}^{-1})-{\cal H}
(z,\bar{z})\ ,
\ee
and
\add
\renewcommand{\theequation}{\arabic{equation}c}
\be
L=\frac{i\hbar}{2}<K^3_q-k>(\dot{\xi}\xi^{-1}-\dotxi\bar{\xi}^{-1})-{\cal H}
(\xi,\bar{\xi})\ .
\ee
Evaluating explicitly the expectation values in the Langrangians,
we find that they are modified with respect to their $q=1$ value [16-19],
due to the $q$-deformation.
Let us
recall further the fact that in the non-deformed cases $\alpha,z$ and
$\xi$ are the coordinates [20] of the respective cosets (generalized phase
spaces):  $\frac{WH}{U(1)}\approx R^{2},\ \frac{SU(2)}{U(1)}\approx S^1
\approx CP^1$ and
$\frac{SU(1,1)}{U(1)}\approx S^{1,1}\approx CP^{1,1}$ and that these
spaces are K\"ahler manifolds with respective potentials $\Phi=\ell n(\alpha|
\alpha)$,
$ \frac{1}{2j}\ell n(z|z)$ and $\frac{1}{2k}\ell n(\xi|\xi)$,
where the states are the usual coherent states of
these groups and $j$ and $k$ are the Casimir and Bargmann indices labelling
the representations. This potential provides after exterior differentiation
the canonical 1-form and the invariant metric, $ds^2=
\partial_A\partial_{\bar{A}}\Phi dAd\bar{A}$ together with the symplectic
2-form $\omega=\frac
{i}{2}
\partial_A\partial_{\bar{A}}\Phi dA\wedge d\bar{A}$, where $\partial_A\equiv
\partial/\partial_A$ [8]. In the presence of deformation however the
involvement
of $q$-CS, changes the K\"ahler potential which in turn changes the metric
distance and the symplectic structure of the phase space.
We find that
\renewcommand{\theequation}{\arabic{equation}a}
\be
ds^2=(<T^-_iT^+_i>-<T^-_i><T^+_i>)dAd\bar{A}
\ee
and
\add
\renewcommand{\theequation}{\arabic{equation}b}
\be
\omega_i=\frac{i}{2}(<T^-_iT^+_i>-<T^-_i><T^+_i>)dA\wedge d\bar{A}.
\ee
The explicit evaluation of the metrics for arbitrary $q$ deformation parameter
will be postponed until a future study and here we will continue by
observing that
on physical grounds one would expect
that the physical mechanism of deformation, albeit elusive at
present, has
quantitatively  a rather perturbative character on the non-deformed models on
which it would  applied.
Reasoning in this way we will proceed by expanding all the involved operators
and states to powers of $\gamma$ ($q=e^\gamma$) and keep only the first order
terms.

Using the expansion
$$[x]=x+\frac{\gamma^2}{6}(x-x^3)+\frac{\gamma^4}{360}(7x-10x^3+3x^5)+\cdots$$
and the deforming maps connecting the deformed generators of our algebras with
their non-deformed counterparts [21] we can express the CS generating
operators in power series of the deformation parameter as follows:
\renewcommand{\theequation}{\arabic{equation}a}
\be
T^+_a=a^++\frac{\gamma^2}{12}\biggl\{(1-N^2)a^+\biggr\}+\cdots\ ,
\ee
and
\add
\renewcommand{\theequation}{\arabic{equation}b}
\be
T^+_J=J^+-\frac{\gamma^2}{12}\biggl\{(2J^3+2j-1)J^+\biggr\}+\cdots\ ,
\ee
and
\add
\renewcommand{\theequation}{\arabic{equation}c}
\be
T^+_K=K^++\frac{\gamma^2}{12}\biggl\{(2K^3+2k+1)K^+\biggr\}+\cdots\ ,
\ee
where $a^+,J^+$ and $K^+$ etc, without sub-$q$ are ordinary step operators
of the respective algebras.

Next we employ the Backer-Champbell-Hausdorff (BCH) formula [22],
$$\exp(A+B)=\exp A\exp B\exp C_2 \exp C_3\cdots$$
with
\begin{eqnarray}
& &C_2=-\frac{1}{2}[A,B]\nonumber\\
& &C_3=\frac{1}{6}[A,[A,B]]+\frac{1}{3}[B,[A,B]]\hspace{1.0cm}{\rm etc}\
,\nonum
\end{eqnarray}
which is valid for any two non-commutative operators, to provide to first order
the following relations between (un-normalized) $q$-CS, $|\alpha)_q,\
|z)_q,\ |\xi)_q$ and their respective non-deformed ones $|\alpha),\
|z)$ and $|\xi)$:
\renewcommand{\theequation}{\arabic{equation}a}
\be
|\alpha)_q\cong|\alpha)-\frac{\gamma^2}{12}\biggl\{\alpha a^+(N-\alpha a^+)
(N-\alpha a^++2)+\nonumber\\
\frac{1}{2}\alpha^2a^{+2}(2N-2\alpha a^++3)+\frac{1}{3}
a^3a^{+3}\biggr\}|\alpha)\ ,
\ee
and
\add
\renewcommand{\theequation}{\arabic{equation}b}
\be
|z)_q\cong |z)-\frac{\gamma^2}{12}\biggl\{
2zJ^3J^+-z^2J^{+2}+(2j-1)zJ^+\biggr\}
|z)\ ,
\ee
and
\add
\renewcommand{\theequation}{\arabic{equation}c}
\be
|\xi)_q\cong|\xi)-\frac{\gamma^2}{12}\biggl\{ 2\xi K^3K^++\xi^2K^{+2}+(2k+1)
\xi K^+\biggr\}|\xi)\ .
\ee
 Utilizing now these expansions we can calculate the K\"ahler potentials
$\Phi_q$ and the metrics for each case. The symplectic 2-forms
to first order in the parameter $\gamma$ are the following for the three
algebras of our study:
\renewcommand{\theequation}{\arabic{equation}a}
\be
\omega_a=\biggl\{ 1-\frac{\gamma^2}{2}\ |\alpha|^2(|\alpha|^2+2)\biggr\}
d\bar{\alpha}\wedge d\alpha\ ,
\ee
and
\add
\renewcommand{\theequation}{\arabic{equation}b}
\be
\omega_\ell&=&\biggl\{ i\ell p^{-2}-\frac{\gamma^2}{6}i\ell\biggl\{ 2\ell
p^{-1}
(\mp 14\ell-4)\tau p^{-2}+\nonumber\\
& &\hspace{2.0cm}+((8\ell\pm 3)\tau^2+(\pm 12\ell+6)\tau)p^{-3}\\
& &\hspace{2.0cm}+(\pm 2\ell+1)(\mp 10\tau^2-5\tau\mp 1)p^{-4}\biggr\}
d\bar{\theta}\wedge d\theta ;\nonumber
\ee
where the upper sign corresponds to the $SU_q(2)$ case and the lower sign
to the $SU_q(1,1)$, and in correspondance $\ell=j$ or $k$ and $\tau=|z|^2$ or
$|\xi|^2$ and $p=1+|z|^2$ or $1-|\xi|^2$ when $\theta=z$ or $\xi$.
Similarly the distance metric can be read from the above formulae by
dropping the wedge products in the r.h.s. As was mentioned in the beginning
the deformation manifests itself geometrically in the cosets of the
groups, and moreover now we also note that the $\gamma^2$ proportional terms
are depended on the modulo of the projective coordinates, which implies an
invariance under phase changes of the complex coordinates, for the additional
terms in the metrics induced by the deformation.

As a further probe to the geometrical effects of the deformation we will
take up the evaluation of the curvature scalar which for our Riemann metric
derived from the $\Phi_q$ K\"ahler potential given by the overlap of
$q$-CS, is taken the form,
$$R=-(\partial_A\partial_{\bar{A}}\Phi_q)^{-1}\partial_A\partial_{\bar{A}}
(\ell n \partial_A\partial_{\bar{A}}\Phi_q)\ .$$
For the case of the $q$-oscillaator we find
$$R=\gamma^212(1+2|\alpha|^2)\ ,$$
and similar results hold for the two other cases. Obviously there is a
non-zero position depended curvature with rotational symmetry which tends
to zero approaching the zero deformation limit.

Before closing some remarks are in order; the way the K\"ahler potentials
and the derivable from them  Riemannian and symplectic metrics, were introduced
above, was all by analogy with the non-deformed case. In the non-deformed
case however these
metrics are  invariant and covariant correspondingly under the respective
canonical transformations associated with each algebra. These same canonical
transformations will not  however possess the right covariance  properties
when applied to the metrics derived above, due to the
modifications of latter by extra  deformation terms
(considering first order deformation chances for
simplicity). Generalized canonical transformations appropriate for the above
deformed metrics are therefore required and we hope to take up this problem
elsewhere.

In conclussion, a geometrical understanding of the $q$-deformation of
the oscillator, $SU(2)$ and $SU(1,1)$ algebras has be searched here
using the $q$-CS path integrals.
It is interesting to note that the association of curvature with
non-co-commutative in the quantum group level has been discussed before
[23] and could probably be related to the curvature found here in the
quantum algebra level, by the existing duality between quantum
algebras and groups as Hopf algebras. Potential applications of the
present results would include the use of quantum groups in
addressing quantum mechanical problems in spaces of non-constant
curvature, and some problems of this kind are now under study.
\vskip 1.0cm
\ \  I should like to thank M.Chaichian, J.R.Klauder and P.P.Kulish
for discussions and the anonymous referee for suggestions leading to the
improvement of the paper.I also wish to thank DGICYT (Spain) for financial
support.
\pagebreak

\end{document}